# A *ROSAT* SEARCH FOR X-RAY EMISSION FROM QUASAR HOST CLUSTERS

Patrick B. Hall

Steward Observatory, University of Arizona, Tucson, AZ 85721

E-Mail: pathall@as.arizona.edu

Erica Ellingson

Center for Astrophysics and Space Astronomy, CB 391, University of Colorado, Boulder, CO 80309

Richard F. Green

National Optical Astronomy Observatories, Tucson, AZ 85719

and

Howard K. C. Yee

Department of Astronomy, University of Toronto, Toronto, Ont., Canada M5S 1A7

## ABSTRACT

We report the results of a search for X-ray emission from quasar host clusters at moderate redshift using the *ROSAT* HRI. We detect no emission from the host clusters of 3C 263 (z=0.646) and PKS 2352-34 (z=0.706) to $3\sigma$ limits of 3.26 and 2.86 $\times 10^{44}$ ergs s$^{-1}$ respectively ($H_o$=50, $q_o$=1/2) for clusters with $r_{core}$=125 kpc and T=5 keV. These limits show that these quasar host clusters are not substantially more X-ray luminous than optically or X-ray selected clusters of similar richnesses at $z \gtrsim 0.5$. We also report the possible detection of a clump of X-ray emitting gas coincident with the brightest radio lobe of 3C 263. This may be evidence for the existence of a clumpy ICM in the host cluster of 3C 263.

*Subject headings:* Clusters of Galaxies, Quasi-Stellar Objects

## 1. Introduction

The association of quasars with galaxies at similar redshifts allows one to use quasars as signposts for locating galaxies and galaxy clusters at high redshift. This provides an efficient way to form samples of galaxies and galaxy clusters for the study of the evolution of such objects. A significant fraction of radio-loud quasars with $0.2<z<0.7$ are known to be situated in Abell richness 0–1 clusters of galaxies (e.g., Yee & Green 1987 (YG87), Ellingson et al. 1991 (EYG91), and Ellingson & Yee 1994). In addition, quasars located in such rich clusters are seen to evolve 5–6 times faster than their counterparts in poor environments (Ellingson et al. 1991, Yee & Ellingson 1993), evolution which may be extrapolated to include the very faint AGN activity seen in radio galaxies in rich clusters (e.g. DeRobertis & Yee 1990). One scenario which can explain these observations is that the physical conditions in the cores of clusters containing optically bright quasars have undergone substantial changes on roughly a dynamical time scale, causing the subsidence of the quasar activity. By combining X-ray images with optical and radio imaging and optical spectroscopy of these clusters, and by comparing quasar host clusters with clusters not harboring quasars, it should be possible to discriminate between different scenarios for this evolution and the role played by the intra-cluster medium (ICM) in it. Comparison of the X-ray properties of quasar host clusters and clusters at similar redshifts discovered using other techniques also provides a broader look at the evolution and diversity of the properties of clusters and their constituent galaxies.

Even though it is thus well established that environment plays a strong role in the triggering and evolution of quasar activity, the physical processes responsible for this link are unclear. Many possible mechanisms have been proposed, including galaxy-galaxy interactions and mergers (Hutchings et al. 1984), cooling flows (Fabian et al. 1986), and an ICM of significantly different density at higher redshifts (Stocke & Perrenod 1981, Barthel & Miley 1988). These different models have considerably different implications for X-ray observations of quasar host clusters.

The interaction/merger model predicts that clusters containing quasars are less evolved dynamically, are still collapsing, and are not virialized (e.g. Roos 1985). Spectroscopy of a large number of galaxies associated with quasars yields cluster velocity dispersions which are significantly lower than for normal Abell clusters, in support of this model (EYG91). Such younger clusters might be expected to have weaker, more irregular X-ray emission than is typical at their redshift (Jones & Forman 1984) However, quasar host clusters sometimes have very high core galaxy densities (Ellingson & Yee 1995). This is consistent with a high merger rate if the clusters are still dynamically young, but for low redshift X-ray clusters high core density is normally associated with the regular X-ray morphology of evolved clusters (Jones & Forman 1984). Thus, clusters discovered by their association with quasars may form a distinct morphological group, complicating the prediction of what their X-ray emission should look like.

Cooling flows are another mechanism for fueling quasars which predicts their existence in the centers of rich clusters (Fabian et al. 1986). With an X-ray luminous object located at the center of the cluster, it is difficult to detect the primary signature of cooling flows – highly peaked X-ray emission – in an X-ray image. Still, this model predicts that quasar host clusters should have regular X-ray morphologies and high core densities, in probable contrast to the interaction/merger model.

A third model was suggested by YG87 and EYG91 based on the model of Stocke and Perrenod (1981). A low density ICM ($\lesssim 10^{-4}$ cm$^{-3}$) might be more amenable to quasar activity than a high density ICM, since stripping is less effective and a more substantial amount of interstellar gas may remain within the host galaxy as a supply of fuel. Stripping of galaxy halos might also decrease the effectiveness of mergers as a quasar fueling mechanism, thus contributing to the decline of quasar activity. There are two pieces of circumstantial evidence that suggest that quasar host clusters have a low-density ICM. First, the brightest quasars found in clusters often have double-lobed FRII morphologies (Yee et al. 1989) which at low redshift are almost always seen only in poor, low ICM density environments. Using this scenario to link clusters hosting bright high-redshift quasars and fainter low-redshift AGN implies that the ICM density has increased since z∼0.5, possibly causing a decline in quasar activity in clusters. Second, many high-redshift quasar host clusters contain a significant fraction of blue galaxies, especially in the cluster core (Ellingson & Yee 1995). This suggests that stripping may not have reached an efficient stage in these clusters, consistent with the existence of a low density ICM. This model can be tested by estimating the ICM densities of quasar host clusters from measurements of their X-ray emission (cf. Arnaud 1988).

Alternatively, Barthel & Miley (1988) suggest that the ICM (or IGM) around z>1.5 radio-loud quasars was clumpy and up to ∼$10^2$ times denser than it is at low redshift, based on the observed lack among such quasars of radio lobes with sizes comparable to those seen at lower redshift. Our quasars are only at z∼0.7, but estimates of their host clusters' ICM densities would be useful in determining the rate of evolution of the ICM density, in conjunction with this or other models.

X-ray images of quasar host clusters are also useful for the study of the evolution of galaxy clusters themselves, since high-redshift clusters show significant evolution in both optical and X-ray properties (e.g. Butcher & Oemler 1984, Gioia et al. 1990, and others). Optical studies of quasar host clusters already show that they are significantly different from low-redshift clusters both dynamically and in the gas content of their member galaxies. Rich clusters at similar redshifts discovered by optical surveys (e.g. Gunn et al. 1986) or especially by X-ray selection (e.g. Gioia & Luppino 1994) probably tend to be evolved clusters, whereas quasar host clusters may tend to be young, and thus may serve as the best sample for discovering evolutionary signatures in clusters of galaxies.

Thus, in order to study the X-ray properties of quasar host clusters, we proposed to use the *ROSAT* High Resolution Imager (HRI; Zombeck et al. 1990) to image half a dozen radio-loud quasars known to lie in rich clusters. We present here the results of the useful observations of two quasars we have obtained so far. We discuss the implications of our results in the light of other recent X-ray observations of high-redshift clusters and the possibilities for advancing our understanding of quasar host cluster properties with future X-ray observations.

## 2. Observations

Table 1: Summary of Objects and Observations

| Name | z | RA (1950.0) | Dec (1950.0) | $N_H^a$ | Dates Observed | Exposure[b] | Quasar Flux[c] |
|---|---|---|---|---|---|---|---|
| PHL 658 | 0.450 | 00:03:25.07 | +15:53:07.4 | 3.95 | 1/7-1/9/1992 | 4737.44 | 1.37 |
| 3C 263 | 0.646 | 11:37:09.34 | +66:04:26.9 | 0.82 | 11/4-11/7/1991 | 26036.41 | 0.38 |
|  |  |  |  |  | 4/18-4/21/1993 | 31192.98 | 0.61 |
| PKS 2352-34 | 0.706 | 23:52:50.62 | −34:14:39.5 | 1.07 | 5/18-6/13/1993 | 40173.94 | 0.29 |

[a]Units of $10^{20}$ cm$^2$; values taken from Elvis et al. 1989 and Lockman & Savage 1994.
[b]"Live time" in units of seconds.
[c]Units of $10^{-11}$ ergs cm$^{-2}$ s$^{-1}$.

Four quasars were scheduled for observations for this program. Unfortunately, our highest-priority targets were not among them; we observed brighter, higher-redshift quasars than we had deemed optimum. Table 1 contains information on the three objects which were actually observed–an observation of 3C 206 was scheduled but no data were obtained due to spacecraft malfunction. PHL 658 was observed successfully, but the exposure time was not long enough to be useful for this project, so that data will not be discussed further. 3C 263, whose host cluster has been spectroscopically confirmed to be Abell richness class $\gtrsim 1$, was observed at two distinct epochs about eighteen months apart, between which its flux increased by about 50%. The first observation was processed with SASS 5_6, and thus may contain small errors in the timing of photon events. These will not affect our analysis. PKS 2352-34, which is surrounded by a galaxy excess consistent with an Abell richness 1 host cluster that, however, has not been spectroscopically confirmed, was observed just once. The quasar fluxes listed are computed from the counts in a circle of radius 1′, with the background from 12-15′ subtracted off. These counts were corrected by 6% for the emission expected to lie outside a circle of that size, given the Point Response Function (PRF) of the HRI. The count rates were converted to fluxes via the procedure detailed in the next section using conversion factors from David et al. (1992) assuming a power-law spectrum with energy index $\alpha$=2.0. There was no evidence for significant flux variability of any of the quasars during the individual pointings, on timescales of a day or so. However, the flux of 3C 263 increased by about 50% between the two observations, spaced a year and a half apart, that we obtained of it.

For an initial look at the data, we smoothed the raw images with a gaussian of $\sigma$=4 pixels (2″). This smoothing size was the smallest which ensured the quasars had smooth radial profiles which decreased monotonically outwards. Smoothed contour plots of all three observations are shown in Figure 1. None of the quasars show obvious extended structure in the smoothed images, although all of them are elongated to some degree. It is known that residual errors in the spacecraft's aspect solution can cause asymmetries in images of point sources (David et al. 1992). Some way of estimating these aspect errors is needed to determine for certain how much, if any, of the observed ellipticity is intrinsic. None of our fields have a bright point source other than the quasar to compare with, but we do have two independent images of 3C 263 to compare. The position angle of the emission differs noticeably, by 30°, between the two observations, and thus the elongation is not likely to be intrinsic to the source.

We experimented with many different methods to correct for the aspect errors to increase our sensitivity to extended emission near the quasars. For example, we looked for differences in the pulse-height spectra of the two observations of 3C 263, since the HRI has a limited energy resolution capability (David et al. 1992). If a cluster contributed substantially to the measured flux from 3C 263, we would expect the spectum to be softer when the flux from the quasar is lower. However, the pulse-height spectra of our two 3C 263 observations are not significantly different, and what difference there is works in the opposite sense from expected.

We also investigated the techniques for correcting for aspect errors in $ROSAT$ HRI data presented by Morse (1994). The first technique he suggests is binning the final image back into the original 'observation intervals', which are the continuous individual exposures that are summed to construct the final image in the standard processing procedure. In each of these 'subimages', which consist of data from only one observation interval, the quasar's centroid is found, and then they are offset to a common centroid and coadded to form a new image. We experimented with this correction, but found that the offsets were in almost all cases within the uncertainties on the mean centroid, and so using them produced only a marginal reduction in the FWHM and ellipticities of the objects. The second technique suggested by Morse assumes that aspect errors are a function of phase in the satellite wobble. The satellite is wobbled solely to prevent occultation of sources in the PSPC field by the thin wire grid supporting the entrance window, but it also functions as a method of flatfielding for the HRI, since as the satellite is wobbled the flux from objects in the field falls onto different parts of the detector. The overall area on the detector containing all the flux from a given object will be a strip if all the observation intervals are wobbled in approximately the same direction, which was the case for our data. By selecting photons from a $1'$ area surrounding the quasar on the final image and binning them into subimages according to where on this strip they were detected, we can look for image centroid offsets as a function of phase in the wobble. The offsets definitely showed a dependence on wobble phase; but again, the amplitude was small enough that only marginal improvements in the FWHM and ellipticities were achieved once a correction was made for this effect. However, this may be due primarily to the limited number of bins we were able to use for the correction, given the small number of object photons detected, rather than demonstrating a dependence of aspect errors on variables other than wobble phase. In addition, not all photons originally detected from the quasar fall within the bins around the strip on the detector used to construct this corrected image. Thus, the signal-to-noise ratio is lowered, which results in the corrected image being worse for calculating upper limits than the original image. In any case, with both these corrections applied, we still see no obvious extended emission around the quasars in either the corrected images or radial profiles.

We looked to see if the X-ray position angle was similar to the position angle of the radio axis. We could not locate a radio map of PKS 2352-34 in the literature, but maps of PHL 658 (Miller et al. 1993) and 3C 263 (Leahy et al. 1989) show that they are both double-lobed FRII sources.

Miley & Hartsuijker (1978) list a position angle of 115±6° (measured from the N, increasing to the E) for the axis of PHL 658, but the position angle of the X-ray emission is 149±3°. 3C 263 is a triple with the brightest component (the eastern lobe) at a displacement of +14.9″,–5.9″ from the core (Owen et al. 1978), equivalent to 16″ away at a position angle of ∼115°±12°. The X-ray emission position angle is quite different from this: 80°±3° in the first observation and 50°±3° in the second. Thus, in the absence of evidence otherwise, we assume that the ellipticity observed in our objects is entirely due to aspect errors.

However, a faint clump of emission is visible in both the 3C 263 images in approximately the same location as the brightest radio component (Fig 1b,c). The clump is displaced by 13.5±0.5″,–6.5±0.5″ from the emission peak, equivalent to 15″ away at a PA of 116°±3°. This clump is quite faint, but it is brighter than other faint "structure" which does not appear in both images. Within a 3″ radius circle centered on the peak of emission from the clump there are 9 counts in the coadded image, where 1.75±0.37 are expected, based on the background measured in a 6″ wide annulus with inner edge 12″ from the quasar, excluding pixels within 6″ of the clump. Optical images show no galaxies coincident with the clump, but there are two faint (i=21.3 and 22.7) galaxies 2.6″ and 3.2″ away from it. The clump has 7.25 background-subtracted counts in a 3″ circle, which should contain 40% of the emission for an unresolved source, given the resolution of the *ROSAT* HRI. Thus we estimate a total of 18.1 counts in our image are from the clump, for a total luminosity of $L_X$=3.9±1.3×$10^{43}$ ergs s$^{-1}$ if it is at the distance of 3C 263 ($H_o$=50, $q_o$=0.5), where $L_X$ is the luminosity in the *observed ROSAT* passband of 0.1-2.4 keV. Given the faintness of the galaxies near the clump, we would expect them to have only $L_X$<$10^{42}$ ergs s$^{-1}$ from the $L_X$–$L_B$ correlation (Fabianno 1994) if they are at the quasar redshift. Thus it is unlikely that the emission, if real, is due to one or both of the galaxies located near the clump's position.

With so few counts attributable to the clump, it is difficult to assign a significance to this 'detection' without doing, for example, a Monte Carlo simulation. However, since the emission appeared in both X-ray images at the same location within the errors, and since its position is coincident within the errors with the eastern radio lobe of the quasar, we believe the emission may be real. (The emission was also detectable in both images after corrections were made to reduce aspect errors, as detailed earlier in this section. We also note that there is no evidence for any excess emission at the site of the western radio lobe, which is about twice as far away from the quasar as the eastern lobe and is four times fainter.) If the emission is real, we are likely seeing an interaction between the quasar's radio jet and the ICM, similar to that seen in great detail in the cluster surrounding Cygnus A (Carilli et al. 1994). The emission may be thermal emission from ICM compressed and displaced outward by the jet, which might indicate an ICM less dense at these redshifts than at z=0, detectable in our images only where it is compressed and heated by the radio jet. Alternatively, the radio jet may be interacting with only a clump of ICM material (recall that the western lobe is located further from the quasar), which would make this observation evidence for a clumpier ICM in high-redshift clusters, as proposed by Barthel & Miley (1988). Spectral information would help decide between these possiblities, but the faintness of the clump and the limited spectral response of the HRI makes such information unobtainable from our data.

## 3. Analysis

Since there was no obvious extended emission in any of our images, we turned to more sophisticated ways of detecting or setting the strongest limits possible on such emission from our data. Note that in our analysis we have assumed that the center of any quasar host cluster emission is coincident with the quasar. If that is not the case, we might expect to see a clear off-center peak in the X-ray images, which we do not see. In any case, the analysis described here could be repeated for an off-center cluster and should result in upper limits comparable to or stronger than those presented here for centered clusters.

The best method to quantify the strength of (or put limits on) extended cluster emission would be to measure the *ROSAT* HRI Point Response Function (PRF) from an unresolved on-axis object in the same field of view as the quasar, and to compare the radial profile of the quasar to that PRF. Unfortunately, neither of our fields has a bright pointlike X-ray source besides the quasar within them. Although at first there appeared to be such an object in the PKS 2352-34 field, inspection of a POSS plate showed that the optical counterpart of the source is a galaxy.

The next best method for searching for cluster emission is to compare the quasar profile with the standard *ROSAT* HRI PRF, which has been well characterized via long exposures of bright, pointlike Galactic X-ray sources (David et al. 1992). As long as the PRF of our observations did not deviate significantly from the norm, we can subtract off the standard PRF and look for excess emission which might be due to a cluster. In the absence of such emission, we can compare the upper limits determined for each annular bin to the expected emission profiles for different clusters, and determine how sensitive we are to each type of cluster.

We did not apply the corrections suggested by Morse (1994) to our data before subtracting the PRF for several reasons. The corrections do decrease the FHWMs and ellipticities of the quasar images slightly, but not all the photons detected can be corrected by this method, and so the S/N is degraded due to the lost flux. In addition, our analysis can require the measurement of flux up to 30″ away from the quasars, which is not corrected by the Morse method. Tests using corrected images showed our limits would not improve if the corrections were applied, so we opted not to apply them.

To do our modelling, we needed object, PRF, and cluster images. We created IRAF[1] images of our objects from the *ROSAT* QPOE observation files at 2x2 binning, thus giving 1″ pixels.

---

[1] The Image Reduction and Analysis Facility (IRAF) is distributed by the National Optical Astronomy Observatories, which is operated by the Association of Universities for Research in Astronomy, Inc., under contract to the National Science Foundation.

These images were not smoothed for this analysis. To measure the flux in circular annuli around the quasar, we created a set of pixel mask files using the IRAF/PROS[2] task *plcreate*, blocking out an r=22.5″ circle around all objects detected by the standard processing, and using annuli of widths from 2.5-60″. For 3C 263, we performed our analysis on each image individually, as well as on a coadded radial profile. We used *immodel* to generate images of the standard *ROSAT* HRI PRF: an exponential halo with scale 31.69″ and normalization 0.0008, and two Gaussians of $\sigma$=2.19″ and 4.04″ and normalizations 0.8421 and 0.1571 respectively (David et al. 1992). (Note that there is an additional PRF component that becomes important beyond 2′. Even though the normalization of this component is estimated to be 10-100 times smaller than the halo, and the signal from our objects is below the background beyond 2′, we still want to be wary of results dependent on data from $\gtrsim 2'$, as well from $\lesssim 20''$ or so due to asymmetries from residual aspect errors.) We compared the radial profiles of the standard PRF to our object profiles, using both circular annuli and elliptical annuli fit to the object images, and concluded that the standard HRI PRF was an adequate match. There is some evidence that the PRF during some observations was a bit narrower than the typical PRF (see Fig. 2a and 2c), but we chose to use the standard PRF as a conservative assumption. Finally, we used *immodel* to generate images of spherically symmetric clusters for several different core radii and cosmologies, all with standard $\beta$=2/3 King model surface brightness profiles convolved with the PRF using *imsmooth*. We used *imcnts* in conjunction with the various annular masks to annularly bin the counts of all these images.

Using these three sets of radial profiles, we searched for extended emission from each quasar as follows. The background counts and associated uncertainty were estimated from the annulus 12-15′ from the quasar. This annulus will also contain emission from both the quasar and the cluster, but these contributions can be calculated and corrected for in an iterative fashion. This constant background was subtracted from the object's radial profile, and the uncertainty in the background added in quadrature to the overall uncertainty. Next, the PRF was normalized to the background-subtracted object counts in the innermost bin. This innermost bin also contains emission from the cluster, if any is present, but the process can be rapidly improved through iteration, as above. This normalized PRF, which corresponds to the expected intensity profile of the quasar, is then subtracted from the background-subtracted object profile, leaving a radial profile consisting of any excess counts above the typical profile expected for an object of the observed central intensity.

These radial profiles for PKS 2352-34 and 3C 263 are plotted in Figure 2a-c. The dips at r<1′ in Fig. 2a and 2c, which are 2-3$\sigma$ drops, may indicate a narrower than usual PRF, but the upper limits do not change with the use of the narrow-end PRF parameters given in David et al. 1992, nor is any extended emission evident then either. Since no cluster emission is obvious, we derive upper limits from our data as follows. In each annulus, we calculate the 3$\sigma$ upper limit on excess emission from a cluster of a given profile using the $\sigma$ calculated for each annulus. The strongest limit from all the annuli becomes our overall $3\sigma$ limit; for clusters with $r_{core}$=125 kpc, the annulus from 15-30″ (∼75-150 kpc, depending on cosmology) typically gives the strongest limit. We assume there is a cluster whose flux equals the $3\sigma$ limit, sum the flux in all annuli from such a cluster, and divide by the exposure time. This gives us our upper limit (in counts/second within r=15′) for cluster X-ray emission around our quasars.

---

[2]The Post Reduction Off-line Software (PROS) package was developed by the High Energy Astrophysics Division of the Smithsonian Astrophysical Observatory, with assistence from the Space Science Computing Division and the Laboratory for High Energy Astrophysics of the Goddard Scpace Flight Center.

Table 2: Upper Limits on Quasar Host Cluster X-ray Emission

| | 3σ Upper Limit Luminosities[a] | | | | | |
|---|---|---|---|---|---|---|
| Cosmology ($H_o$, $q_o$): | 50, 0.5 | | 75, 0.5 | | 100, 0 | |
| $r_{core}$ (kpc) | 125 | 250 | 125 | 250 | 125 | 250 |
| 3C 263 | 3.26 | 4.78 | 1.74 | 3.05 | 1.42 | 2.47 |
| PKS 2352-34 | 2.86 | 4.10 | 1.54 | 2.54 | 1.21 | 2.01 |

[a]Units of $10^{44}$ erg s$^{-1}$

There are several variables that affect the value of this upper limit: simulated cluster core radius, annulus width, and the cosmology. Thus, for each object, we tested clusters with 2 different core radii (125 & 250 kpc), annular bins of 8 different widths from 2.5–60″, and 3 different cosmologies ($H_o/q_o$ of 50/0.5, 75/0.5, and 100/0.0). The limits we obtain do not depend in a simple fashion on these parameters since the interplay between the variables is quite complex, although they do vary monotonically over the range of interest as each parameter is varied independently. We chose the two core radii values given because EYG found $r_{core}$<150-200 kpc for quasar host clusters and because $r_{core}$=250 kpc is often taken as typical for clusters in general, although values in the range 70-900$h_{50}^{-1}$ kpc were found by Jones & Forman (1984).

Several steps must be taken to convert our limits from counts/second to $L_X$. Using the known value of $N_H$ and assuming a value for kT, we can use the energy-to-counts conversion factors listed in the *ROSAT* HRI Instrument Manual (David et al. 1992). As listed, these factors assume only half the total flux is in the detect cell; so, since we measure the total flux, we multiply the listed factor by 2. We divide this factor into our counts/sec limit to convert to the energy flux F in units of $10^{-11}$ ergs cm$^2$ s$^{-1}$. Next we convert to $L_X$ in the *observed* 0.1-2.4 keV band via the formula F=$L_X/4\pi D_L^2$, where $D_L$ is the luminosity distance for the quasar redshift and assumed cosmology:

$$D_L = \frac{c}{H_o q_o^2 [z q_o + (q_o - 1)(-1 + \sqrt{2 z q_o + 1})]} \quad (1)$$

for $q_o > 0$, and

$$D_L = \frac{cz}{2H_o(z+2)} \quad (2)$$

for $q_o$=0. Finally, we correct the cluster luminosity for emission beyond r=15′, given the cosmology and cluster core radius.

The resulting upper limits, for different cosmologies and core radii, are given in Table 2. For Ho=50, qo=1/2, T=5 keV, and $r_{core}$=125 kpc, we set limits of 3.26 and 2.86 × $10^{44}$ ergs s$^{-1}$ for 3C 263 and PKS 2352-34 respectively. All the limits become roughly 25% lower if we adopt T=1 keV, and 35% for T=0.5 keV, since more of the emission falls in the $ROSAT$ HRI passband at those temperatures than at T=5 keV. The observed T-$L_X$ correlation for X-ray-selected clusters (e.g. Wang & Stocke 1993) yields T$\lesssim$4 keV for our upper limit values of $L_X$, so assuming T=5 keV is a valid conservative assumption.

In the limit as $r_{core}$ approaches zero, cluster emission becomes pointlike, indistinguishable from quasar emission, and our limits should approach the quasar flux. Since our limits were still decreasing at 125 kpc, we simulated a cluster with $r_{core}$=50 kpc to make sure that our limits on emission did not continue to decrease at ever smaller $r_{core}$. And indeed, for a cluster with $r_{core}$=50 kpc our limits are less stringent than for ones with 125 kpc. This raises the question of how sensitive we are to the detection of emission from a clumpy ICM concentrated near the cluster core. Since we may have detected emission from a clump of ICM material with $L_X$=3.9×10$^{43}$ erg s$^{-1}$ at r=15″, we can say that, roughly, any X-ray emitting clumps of the ICM with $L_X \gtrsim$10$^{43}$ erg s$^{-1}$ would have to be located within $\sim$15″ ($\sim$100 kpc) of the quasar to avoid detection due to noise or asymmetries from residual aspect errors.

## 4. Discussion

The launch of $ROSAT$ has enabled the study of X-ray emission from moderate to high redshift clusters selected optically, by X-ray emission, or by hosting radio galaxies or quasars. Here we summarize some such recent results and put our results in context, beginning with studies of optically-selected clusters.

Bower et al. (1994) observed 12 distant ($\bar{z}$=0.42) optically-selected Abell richness class 1–2 clusters with the $ROSAT$ PSPC and found their X-ray emission to be quite weak: the average of the seven detections was $L_{X,44}$=0.44h$_{50}^{-2}$ (where $L_{X,44}$ is the soft X-ray luminosity in units of 10$^{44}$ erg s$^{-1}$), only two of them had $L_{X,44}$>0.5, and none had $L_{X,44}$>1, where $L_{X,44}$ is the X-ray luminosity in units of 10$^{44}$ erg s$^{-1}$. The upper limits on the non-detections are also all <10$^{44}$ ergs s$^{-1}$. Thus, if the present-day correlations between cluster richness and $L_X$ hold at z=0.4, the z=0.4 cluster X-ray luminosity function (XLF) is incompatible at the 3$\sigma$ level with the present-day XLF. This is consistent with the results of earlier studies of high-redshift X-ray selected clusters by Gioia et al. (1990) and Henry et al. (1992).

Castander et al. (1994) observed 5 optically-selected rich clusters with z=0.69–0.92 with the $ROSAT$ PSPC and detected 2 of them, one at z=0.697 with $L_{X,44}$=0.73h$_{50}^{-2}$ and one at z=0.895 with $L_{X,44}$=1.08h$_{50}^{-2}$, along with the serendipitous detection of emission from a cluster at z=0.539 with $L_{X,44}$=1.07h$_{50}^{-2}$. The upper limits on emission from the remaining three are all <10$^{44}$ ergs s$^{-1}$. These luminosities are lower than expected for such rich clusters, based on the correlations observed at low redshift, and so these observations extend the evidence for negative evolution of the cluster XLF to z$\sim$1.

Nichol et al. (1994) presented *ROSAT* PSPC observations of 4 distant cluster candidates, two optically and two radio-source selected. One of each type was detected; the radio-source selected cluster has a photometrically estimated redshift of z=0.68, yielding a value of $L_{X,44}$=2.3$h_{50}^{-2}$, while the optically selected cluster is at z=0.61 based on four redshifts, yielding $L_{X,44}$=1.4$h_{50}^{-2}$. Upper limits for the two non-detected candidates are not calculable since no redshift information is available, but are likely to be $\sim 10^{44}$ ergs s$^{-1}$ assuming the clusters are at z$\gtrsim$0.5. Since the two detected clusters seem to be the most X-ray-luminous non-X-ray-selected clusters at z$\gtrsim$0.5, it would be worthwhile to spectroscopically confirm the cluster redshifts and study the cluster galaxies' properties.

Thus we can see that our limits of $L_{X,44} \lesssim 3$ for emission from the two quasar host clusters are not quite strong enough to determine if their X-ray emission properties are different from optically-selected high-redshift clusters, which have $L_{X,44} \lesssim 1$. A factor of 3 improvement in our upper limits would be enough to confirm that quasar host clusters are not more luminous than the most luminous optically selected clusters at high redshift. Such an improvement should be possible with the observation of less luminous quasars in richer clusters at lower redshift.

As for radio-source selected clusters, Sokoloski et al. (1995) have obtained *ROSAT* HRI images of four such clusters at z$\sim$0.5. For the radio-loud quasar 53W080 (z=0.546, photographic J=18.26), they have a 5.2$\sigma$ detection of emission in a circle of r$\sim$150 kpc around the quasar, corresponding to $L_{X,44}$=0.82$\pm$0.16 ($H_o$=100, $q_o$=0). However, as they point out, the emission appears to be unresolved, and it is likely that much of the detected emission is from the quasar rather than the host cluster ICM. This would bring the cluster emission more in line with the upper limits of $L_{X,44}$=0.06–0.44 they find for the three other clusters in their sample. Still, in either case, the limit on quasar host cluster emission is much tighter than ours and illustrates the benefit of observing less luminous quasars– both 3C 263 and PKS 2352-34 have B$\sim$16.3, roughly two magnitudes brighter than 53W080. Also, Worrall et al. (1994) and Crawford & Fabian (1993) may have detected extended X-ray emission around the radio galaxies 3C 280 (z=0.998) and 3C 356 (z=1.079), respectively, with the *ROSAT* PSPC. For $H_o$=50 and $q_o$=0.5, we estimate that if all the detected emission is from cluster gas, the clusters would have total luminosities of $L_{X,44}$=3.73 and 1.51, respectively. HRI images of these objects would be able to accurately determine how much of the emission is from an extended component. Lastly, Wan et al. (1994) have compiled X-ray data on high-redshift clusters containing FRII radio sources, and find that the clusters with FRII sources tend to be less X-ray luminous than those without. Stronger limits on emission from the host cluster of 3C 263, a FRII source, coupled with more observations of both FRI and FRII quasars and radio galaxies, would help confirm this conclusion.

The Einstein EMSS catalog includes X-ray selected clusters out to z$\sim$0.8 (Gioia & Luppino 1994), but the relatively limited sensitivity of that catalog means that such objects are quite luminous and rich: the 10 most distant EMSS clusters have $\overline{z}$=0.54 and $\overline{L_{X,44}}$=7.7. It is useful to know that such X-ray luminous clusters exist at high redshift, despite their rarity, and future work based on *ROSAT* data should be able to determine the luminosity function of distant X-ray selected clusters down to luminosities comparable to distant optically-selected clusters. Figure 3. shows a plot of $B_{gc}$ vs. $L_X$ for our two clusters (assuming $H_o$=50, $q_o$=1/2, and $r_{core}$=125 kpc) and a moderate-redshift subsample of the X-ray selected EMSS clusters being studied by the CNOC group (c.f. Carlberg et al. 1993). $B_{gc}$ is the amplitude of the galaxy-cluster-center or galaxy-quasar spatial correlation function, and serves as a measure of the richness of the clusters. The dotted line is the best-fit relation to the CNOC data. Although both our objects appear to lie below the general trend followed by the X-ray selected clusters, our upper limits are not strong enough to rule out the possibilty these quasar host clusters do follow the same relation, given the scatter in the CNOC data. A factor of ∼2 improvement in our X-ray upper limits, plus a more accurate measurement of $B_{gc}$ and $r_{core}$ for these two objects, should be able to answer that question.

## 5. Conclusions

We have searched for X-ray emission from two quasar host clusters at moderate redshift using the *ROSAT* HRI. We set $3\sigma$ upper limits on emission from the host clusters of 3C 263 (z=0.646) and PKS 2352-34 (z=0.706) of 3.26 and 2.86 $\times 10^{44}$ ergs s$^{-1}$ respectively ($H_o$=50, $q_o$=1/2, $r_{core}$=125 kpc, T=5 keV). Without positive detections of cluster emission, we cannot set limits to the ICM density in these clusters, which would be an important discriminant between different models of quasar formation and fueling. We do, however, possibly detect emission from a clump of X-ray emitting gas coincident with the brightest radio lobe of 3C 263. This may be evidence for the existence of a clumpy ICM in this cluster.

Our results show that these quasar host clusters are not substantially more X-ray luminous than optically or X-ray selected clusters of similar richness at these redshifts. Improvement in our upper limits by at least a factor of 3 should be possible by using the methods described in this paper on observations with similar integration times on less luminous quasars in richer clusters at similar or lower redshift, especially if residual aspect solution errors can be reduced. Such an improvement would show whether or not quasar host clusters follow the same richness-$L_X$ relation as X-ray or optically selected clusters. In addition, by studying quasar host clusters over a wider range of redshift and absolute luminosity, it would be possible to look for correlations between the evolution of X-ray and optical cluster properties. For example, in the cooling flow model of quasar formation and fueling, we might expect a correlation between quasar and cluster X-ray luminosity, whereas if high ICM density retards quasar activity, there should be an anti-correlation between quasar activity and X-ray emission, assuming the ICM is hot enough to be luminous in the *ROSAT* passband.

Several groups have obtained or have been granted long *ROSAT* HRI exposures on moderate-redshift quasars known or thought to reside in clusters. More limits on, and especially detections of, quasar host cluster X-ray emission from these observations will help extend our understanding of these clusters and reassure us that the two clusters we have examined so far are not unusual.

– 13 –## REFERENCES

Arnaud, K. A. 1988, in *Cooling Flows in Clusters and Galaxies*, ed. A.C. Fabian (Dordrecht: Kluwer)

Barthel, P. D., and Miley, G. K. 1988, Nature 333, 319

Bower, R.G., Boehringer, H., Briel, U.G., Ellis, R.S., Castander, F.J., and Couch, W.J. 1994, MNRAS 268, 345

Butcher, H. R. and Oemler, A. 1984, ApJ 285, 426

Carilli, C.L., Perley, R.A., Harris, D.E. 1994, MNRAS 270, 173

Carlberg, R.G. et al. 1994, JRASC 88, 39

Castander, F.J., Ellis, R.S., Frenk, C.S., Dressler, A., and Gunn, J.E. 1994, ApJ 424, L79

Crawford, C.S., and Fabian, A.C. 1993, MNRAS 260, L15

David, L.P., Harnden, F.R., Jr., Kearns, K.E., and Zombeck, M.V. 1992, The *ROSAT* High Resolution Imager (HRI), Technical Report (US *ROSAT* Science Data Center/SAO)

DeRobertis, M., and Yee, H. K. C. 1990, AJ 100, 84

Ellingson, E., Yee, H. K. C. and Green, R. F. 1991, ApJ 371, 36

Ellingson, E., and Yee, H. K. C. 1994, ApJS 92, 33

Ellingson, E., and Yee, H. K. C. 1995, in preparation

Elvis, M., Lockman, F.J., Wilkes, B.J. 1989, AJ 97, 777

Fabian, A. C., Arnaud, K. A., Nulsen, P. E. J., and Mushotzky, R. F. 1986, ApJ 305, 9

Fabianno, G. 1994, in *New Horizon of X-ray Astronomy: First Results from ASCA*, ed. F. Makino

Gioia, I.M., Henry, J.P., Maccacaro, T., Morris, S.L., Stocke, J.T., and Wolter, A. 1990, ApJ 356, L35

Gioia, I.M., and Luppino, G.A. 1994, ApJS 94, 583

Gunn, J. E., Hoessel, J. G., and Oke, J. B. 1986, ApJ 306, 30

Henry, J.P., Gioia, I.M., Maccacaro, T., Morris, S.L., Stocke, J.T., and Wolter, A. 1992, ApJ 386, 408

Hutchings, J. B., Crampton, D., and Campbell, B. 1984, ApJ 280, 41

Jones, C., and Forman, W., 1984, ApJ 276, 38

Leahy, J.P., Muxlow, W.B., and Stephens, P.W., MNRAS 239, 401

Lockman, F.J., and Savage, B.D. 1995, ApJS 97, 1

Miley, G.K., and Hartsuijker, A.P. 1978, A&AS 34, 129

Miller, P., Rawlings, S., and Saunders, R. 1993, MNRAS 263, 425

Morse, J. A. 1994, PASP 106, 675

Nichol, R.C., Ulmer, M.P., Kron, R.G., Wirth, G.D., Koo, D.C. 1994, preprint

Owen, F.N., Porcas, R.W., and Neff 1978, AJ 83, 1009

Roos, N. 1985, A&A 104, 218

Sokoloski, J.L., Daly, R.A., and Lilly, S.J. 1995, submitted

Stocke, J. T. and Perrenod, S. C. 1981, ApJ 245, 375

Wan, L., Daly, R.A., Jones, L.V., and Lilly, S.J. 1994, in *The Soft X-Ray Cosmos,* AIP Conference Proceedings 313, eds. E.M. Schlegel & R. Petre (New York:AIP Press), p. 386

Wang, Q., and Stocke, J.T., 1993, ApJ 408, 71

Worrall, D.M., Lawrence, C.R., Pearson, T.J., and Readhead, A.C.S. 1994, ApJ 420, L17

Ellingson, E., & Yee, H. K. C. 1993, ApJ 411, 43

Yee, H. K. C. and Green, R. F., 1987, ApJ 319, 28

Yee, H. K. C. , Ellingson, E., Green, R. F. and Pritchet, C. J. 1989, in *The Epoch of Galaxy Formation*, eds. C. S. Frenk et al., (Dordrecht:Kluwer), p. 185

Zombeck, M.V., Conroy, M., Harnden, F.R., Roy, A., Braeuninger, H., Burkert, W., Hasinger, G., and Predehl, P. 1990, in Proc. SPIE Conference on EUV, X-Ray, and Gamma-Ray Instrumentation for Astronomy, ed. O.H.W. Siegmund and H.S. Hudson, Proc. SPIE 1344, p. 267

---

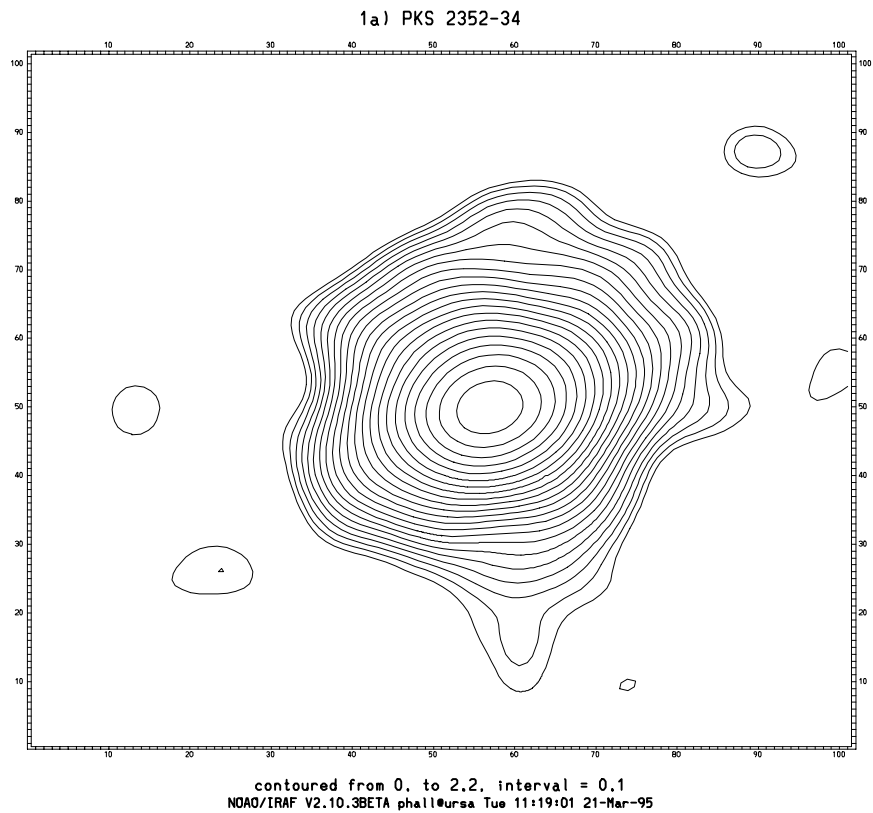

Fig. 1.— Contour plots of all three ROSAT HRI observations. Each image was smoothed with a gaussian of $\sigma=4$ pixels ($2''$) before contouring. A $25''\mathrm{x}25''$ region is shown in each plot; north is up and east is to the left. (a) PKS 2352-34, observed Jan 7-9, 1992. (b) 3C 263 #1, observed November 4-7, 1991. (c) 3C 263 #2, observed April 18-21, 1993.

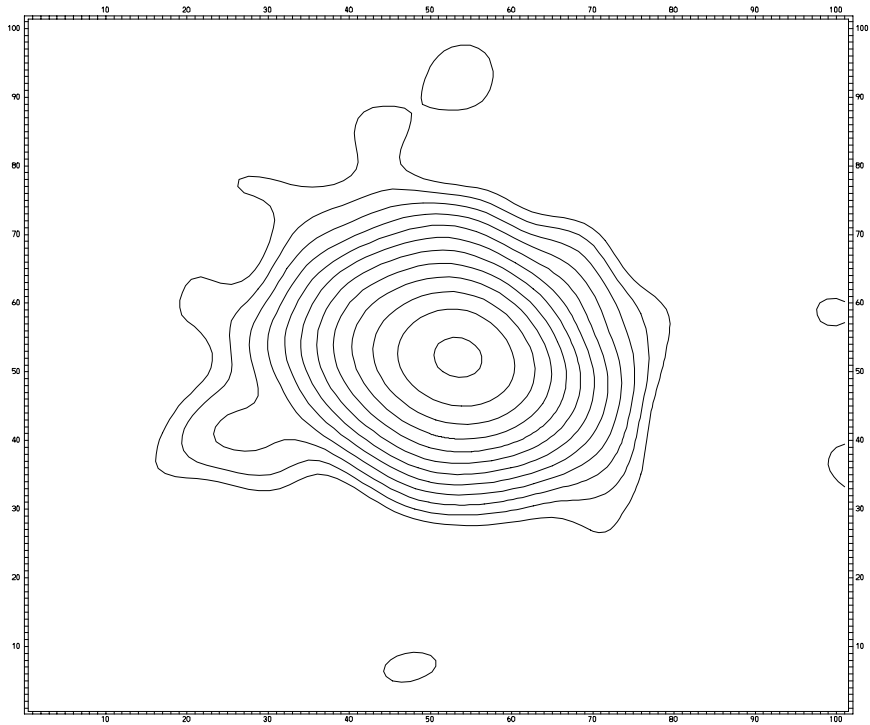

1b) 3C 263 #1

contoured from 0.05 to 2.45, interval = 0.2, labels scaled by 100.
NOAO/IRAF V2.10.3BETA phall@ursa Tue 11:19:51 21-Mar-95

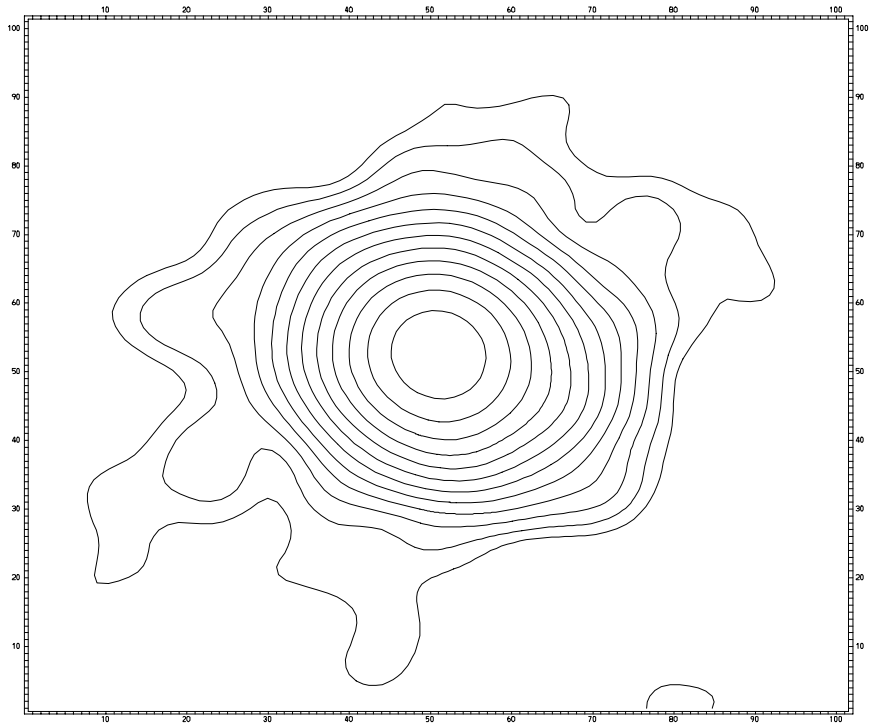

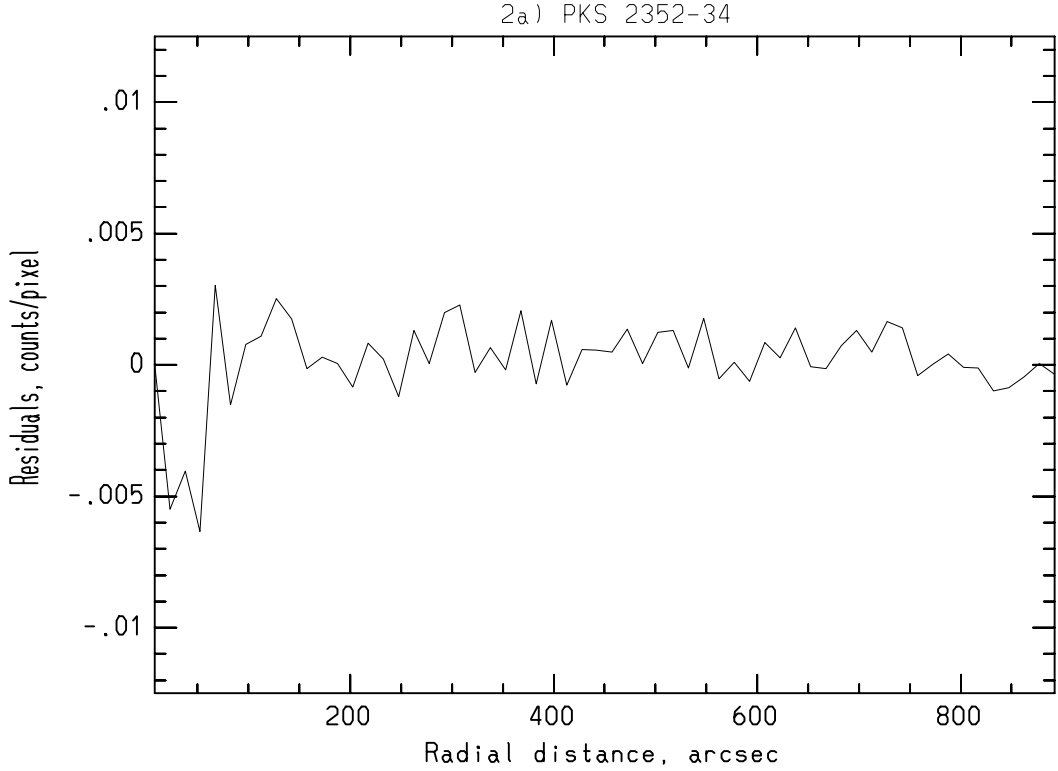

Fig. 2.— Radial profiles of the residuals around the quasar positions after subtraction of the background and of a point response function (PRF) normalized such that the residual is zero in the first bin. (a) PKS 2352-34. The dip at r<1′ a 2-3$\sigma$ drop, may indicate a narrower than usual PRF in the image, but the upper limits do not change with the use of such a PRF. (b) 3C 263 #1. Note the different radial scale of this plot. (c) 3C 263 #2. Again, the dip at r<1′ may indicate a narrower than usual PRF.

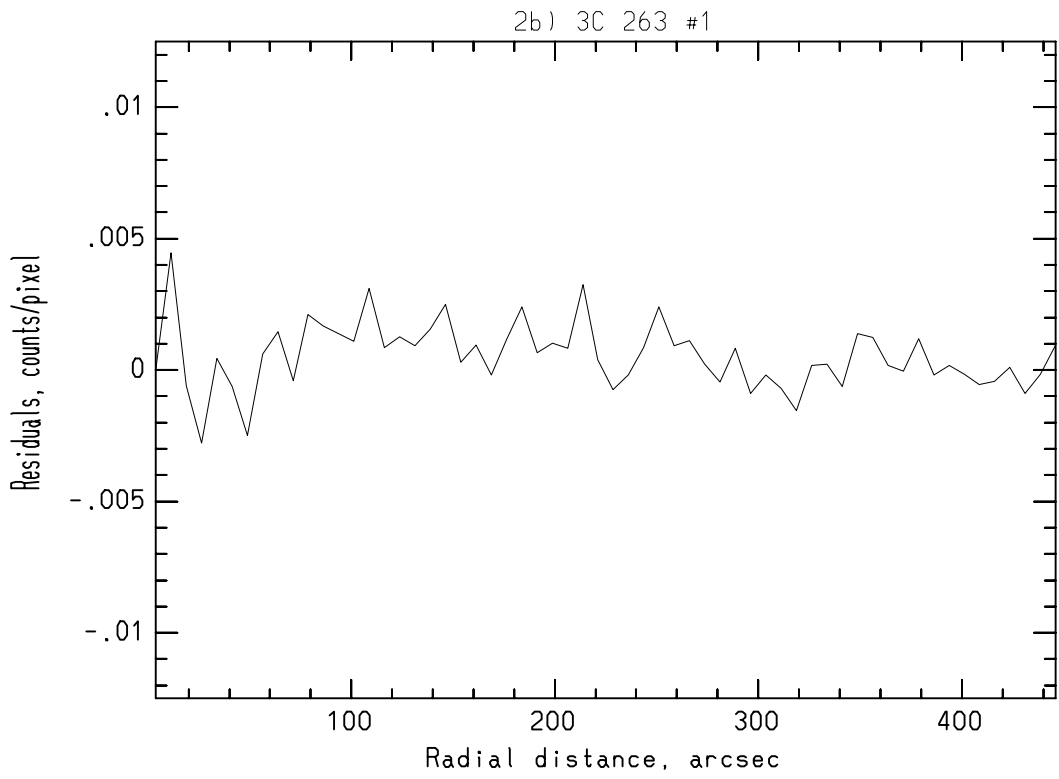

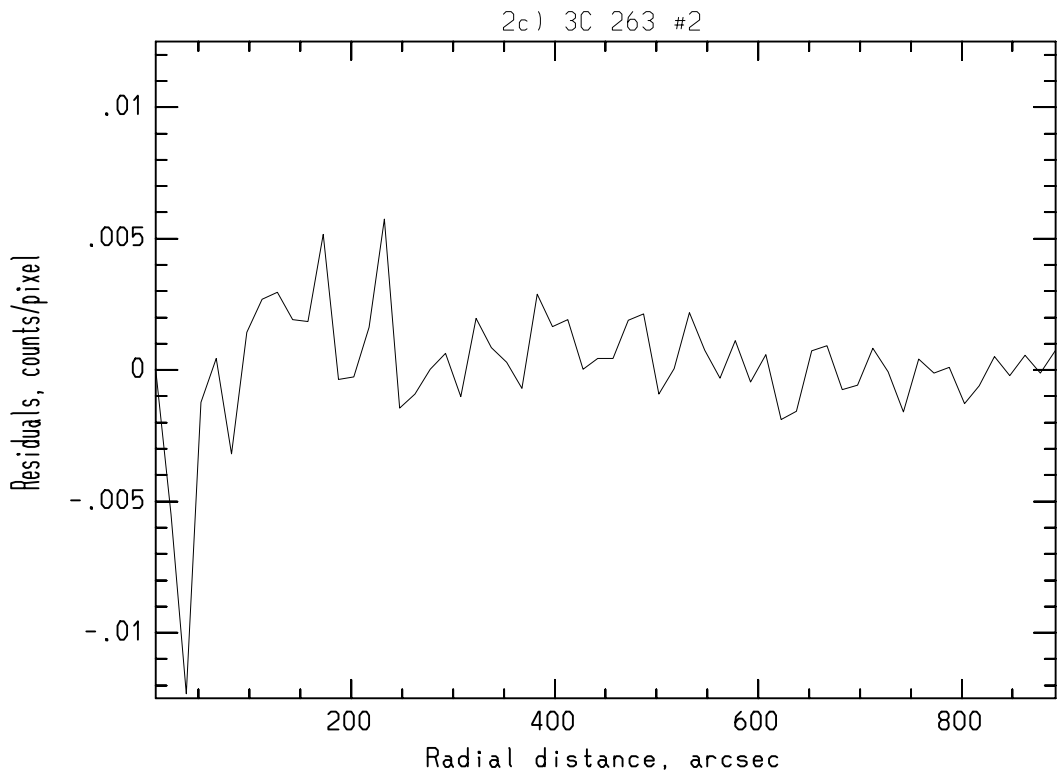

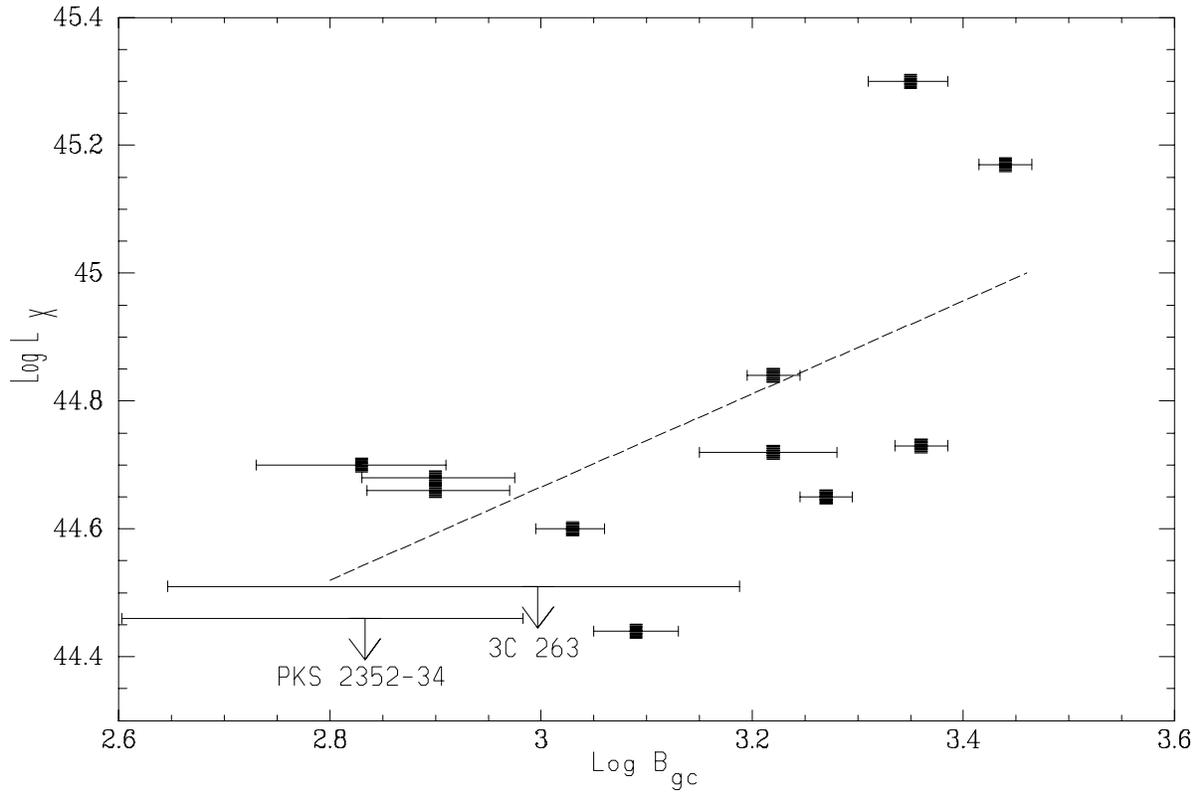

Fig. 3.— A plot of the galaxy-quasar (or galaxy-cluster center) spatial correlation function, $B_{gc}$, vs. the soft X-ray luminosity $L_X$ for our two clusters and a moderate-redshift subsample of EMSS clusters. The dotted line is the best-fit relation to the CNOC data. The upper limits for PKS 2352-34 and 3C 263 assume $H_o=50$, $q_o=1/2$, and $r_{core}=125$ kpc.